# DISSIPATIVE PARTICLE DYNAMICS FOR SYSTEMS WITH POLAR SPECIES

*Alexey A. Gavrilov[§]*

[§]Physics Department, Lomonosov Moscow State University, Moscow 119991, Russian Federation

e-mail: gavrilov@polly.phys.msu.ru

ABSTRACT

In this work we developed a method for simulating polar species in the dissipative particle dynamics (DPD) method. The main idea behind the method is to treat each bead as a dumb-bell, i.e. two sub-beads (the sub-beads can bear charges) kept at a fixed distance, instead of a point-like particle. The interaction forces between such composite beads were studied as well as the relation of the interaction parameters between them to the Flory-Huggins theory; it was shown that at small enough separations the composite beads act essentially as conventional point-like beads. Next, the relation between the bead dipole moment and the bulk dielectric permittivity was obtained. The interaction of single charges in polar liquid showed that the observed dielectric permittivity (i.e. the ratio of the force exerted on the charges in vacuum to the corresponding force in the polar liquid) is somewhat smaller than that obtained for the bulk case at large separation between the charges; at distances comparable to the bead size the solvation shells of the charges start to interfere and oscillations in the observed permittivity occur reflecting the presence of the layers of oriented polar molecules surrounding the charges. The latter are presumably the reason for the observed deviations at large distances as well: such oriented molecules surrounding the charges effectively have smaller polarizability compared to the bulk liquid. Finally, we showed why it is necessary to treat the polar species in DPD explicitly instead of implicitly by calculating the local polarizability based on the local species concentrations: the latter leads to the violation of the Newton's third law resulting in simulation artifacts. We investigated the behavior of a charged colloidal particle at an interface of polar and non-polar liquids. We obtained that when the polar molecules are treated explicitly, the charged colloidal particle moved into the polar liquid since it is energetically more favorable for the charged molecules to be immersed in a polar medium; however, within the "implicit polarity" method the colloidal particle is found on top of a "bump" formed by the molecules of the non-polar liquid, which increases the interface area between the liquids instead of decreasing it.

**Introduction**

In a number of materials, including biological and synthetic, there are phenomena governed by charges in media with non-constant dielectric permittivity. Therefore, the spatial variations of permittivity should be considered in order to treat the electrostatic interactions correctly. This is, however, a rather challenging task due to the emerging theoretical and computational problems associated with that, and the inhomogeneity of the dielectric constant of the medium is often ignored or oversimplified.

The importance of the dielectric constant variations are most pronounced in polymer systems, and the theoretical investigations of the effect of the so-called dielectric mismatch in such systems demonstrate that. The dielectric contrasts of the constituent species in a wide range of polymer systems, such as salt-doped polymer block copolymers, polymer blends, and polymer−ionic liquid mixtures, has been shown to have a dramatic effect on the phase behavior of the system[1–9]; the ion solvation energy can significantly alter the miscibility of a binary polymer blend[10,11]. A recent investigation has shown that dielectric mismatch solely can lead to a new type of microphase separation in polyelectrolyte solutions[12]. The effects of the varying dielectric permittivity cannot be considered using a composition-dependent χ alone as it was shown in a recent work [9] where an asymmetric phase diagram for a salt-doped block copolymer was obtained due to the dielectric mismatch. It is crucial to control the morphology and miscibility of polymer systems as the distribution of ions has been shown to have a dramatic effect on the material properties, for example, ionic conductivity[13,14], so understanding the physics governing the behavior of the systems with varying dielectric constant is of great importance.

Despite the obvious significance of considering the varying medium polarity, only a few simulation techniques for calculating ion distributions in the presence of a varying dielectric constant have been reported [15–19], and most of these approaches have been utilized for investigation of static ion distributions near static dielectric boundaries. Methods allowing for the changes of the dielectric permittivity due to the system evolution in time are also present; the most discussed way is based on the local concentrations of species[18,19]. Such concentrations are further used to assign a local permittivity value to each node of the lattice which is used to calculate the local field. This approach, while being rather effective from the computational point of view, has one disadvantage: there are no electrostatic forces acting on uncharged but polar molecules. This is acceptable for the systems where no redistribution of the species can happen, but can cause unphysical behavior in more complicated cases. Especially that can be problematic for methods where the solvent is explicit such as dissipative particle dynamics (DPD): essentially the interaction caused by the variations in the dielectric permittivity are taken into account in the "implicit solvent" fashion, which violates the Newton's third law. In this work we will discuss another approach to accounting for the polar species by explicitly treating them as dipoles of a fixed length.

**DPD with electrostatic interactions**

As it was mentioned earlier, the main idea of this work is to develop an approach for proper treating polar species in the system within the DPD method. First we give a brief description of the standard dissipative particle dynamics method without electrostatic interactions. Dissipative particle dynamics is a version of the coarse-grained molecular dynamics adapted to polymers and mapped onto the classical lattice Flory–Huggins theory[20–23]. It is a well-known method which has been used to simulate properties of a wide range of polymeric systems, such as single chains in solutions[24], polymer melts[25,26] and networks[27–29]. In short, macromolecules are represented in terms of the bead-and-spring model (each coarse-grained bead usually represents a group of atoms), with beads interacting by a conservative force (repulsion) $F_{ij}^c$, a bond stretching force (only for connected beads) $F_{ij}^b$, a dissipative force (friction) $F_{ij}^d$, and a random force (heat generator) $F_{ij}^r$. The total force is given by:

$$F_i = \sum_{i \neq j} \left( F_{ij}^c + F_{ij}^b + F_{ij}^d + F_{ij}^r \right) \quad (1)$$

The soft core repulsion between $i$- and $j$-th beads is equal to:

$$F_{ij}^c = \begin{cases} a_{\alpha\beta}(1 - r_{ij}/R_c)\mathbf{r}_{ij}/r_{ij}, & r_{ij} \leq R_c \\ 0, & r_{ij} > R_c \end{cases}, \quad (2)$$

where $\mathbf{r}_{ij}$ is the vector between $i$-th and $j$-th bead, $a_{\alpha\beta}$ is the repulsion parameter if the particle $i$ has the type $\alpha$ and the particle $j$ has the type $\beta$ and $R_c$ is the cutoff distance. $R_c$ is basically a free parameter depending on the volume of real atoms each bead represents [23]; $R_c$ is usually taken as the length scale, i.e. $R_c=1$.

If two beads ($i$ and $j$) are connected by a bond, there is also a simple spring force acting on them:

$$F_{ij}^b = -K(r_{ij} - l_0)\frac{\mathbf{r}_{ij}}{r_{ij}}, \quad (3)$$

where $K$ is the bond stiffness and $l_0$ is the equilibrium bond length.

We do not give here a more detailed description of the standard DPD model (without electrostatic interactions); it can be found elsewhere[23].

In order to take into account the electrostatic interactions, we use the method described in the work [30]. The electrostatic force between two charged beads is calculated using the following expression:

$$F_{ij}^e = \frac{q_i q_j}{4\pi\varepsilon\varepsilon_0} \begin{cases} \frac{\mathbf{r}_{ij}}{r_{ij}^3} \sin^6\left(\frac{2\pi r_{ij}}{4D}\right), & r_{ij} < D \\ \frac{\mathbf{r}_{ij}}{r_{ij}^3}, & r_{ij} \geq D \end{cases}, \quad (4)$$

where $D$ is the damping distance. This approach allows one to prevent overlapping of oppositely charged species while keeping the exact form of the Coulomb potential at distances larger than $D$; the parameter $D$ is essential the effective bead size and the electostatic interactions at smaller distances are not important for the system behavior. We used $D=0.65$ which was shown[30] to be a good choice for the number density of 3, which will be used in the present work.

In this work the dimensionless electrostatic coupling parameter[18,30] $\Gamma = \frac{e^2}{4\pi\varepsilon\varepsilon_0 kTR_c}$ was taken equal to 16; we considered similar parametrization to that proposed by Groot[18] but the

"background" medium permittivity ε was taken equal to 5; in what follows, all the relative permittivities will be expressed with respect to this value.

**Dipoles simulation**

The most widely used approach to take into account polar species is to assign a point-like dipole to all the polar beads[31]. Instead of doing that, we will treat each bead as a pair of force centers (we will call them sub-beads in what follows) separated by a small distance *d*. We can assign opposite charges *q* to the sub-beads within one bead, thus making it a dipole with a dipole moment of *qd*. Each sub-bead is treated as a separate force center and interacts with the others through the full set of forces (i.e. eq.1 and 4); the sub-beads positions are integrated independently, which automatically accounts for the dipoles rotation according to the surrounding beads positions and the local electric field. While this approach is less computationally effective compared to the point-like dipoles, the latter has one important disadvantage: when we consider polymer chains with polar monomer units, in the point-like dipole model the dipoles can rotate freely rotate, while in reality the dipole orientations are constrained by the connectivity in chain. The approach proposed in this work takes this into account; moreover, it allows one to consider different polar group location relative to the backbone, i.e. perpendicular, when only one sub-bead of each monomer unit is in the chain backbone, so the dipole can freely rotate around the backbone, or parallel, when both the sub-beads are in the backbone, and the dipoles orientation is firmly related to the backbone local orientation (see fig.1a). Since DPD has been widely used for simulating polymer systems, this feature seems to be rather important for capturing the general system behavior.

The distance between the sub-beads within a bead is kept constant using the standard RATTLE method[32] in order to prevent the beads from changing their size and dipole moment. Essentially, each bead is simulated as a rigid dumb-bell. In order to improve the computation efficiency and allow for using large timesteps (which is one of the DPD main advantages), a somewhat simplified version of RATTLE is used; when solving the quadratic equations to determine the Lagrangian multipliers[31], we neglected the quadratic term since it is proportional to $\Delta t^4$. The resulting linear equations always have a solution; the deviations from the target distance were found to be as small as 1% for the timestep of 0.02 which was used in the present work. The distance *d* between the sub-beads should be small enough so that such a dumb-bell bead structure still has the same properties as a standard point-like bead. Fig.1b shows the force between two dumb-bell beads depending on *d*; this force is compared to the force between two standard point-like beads.

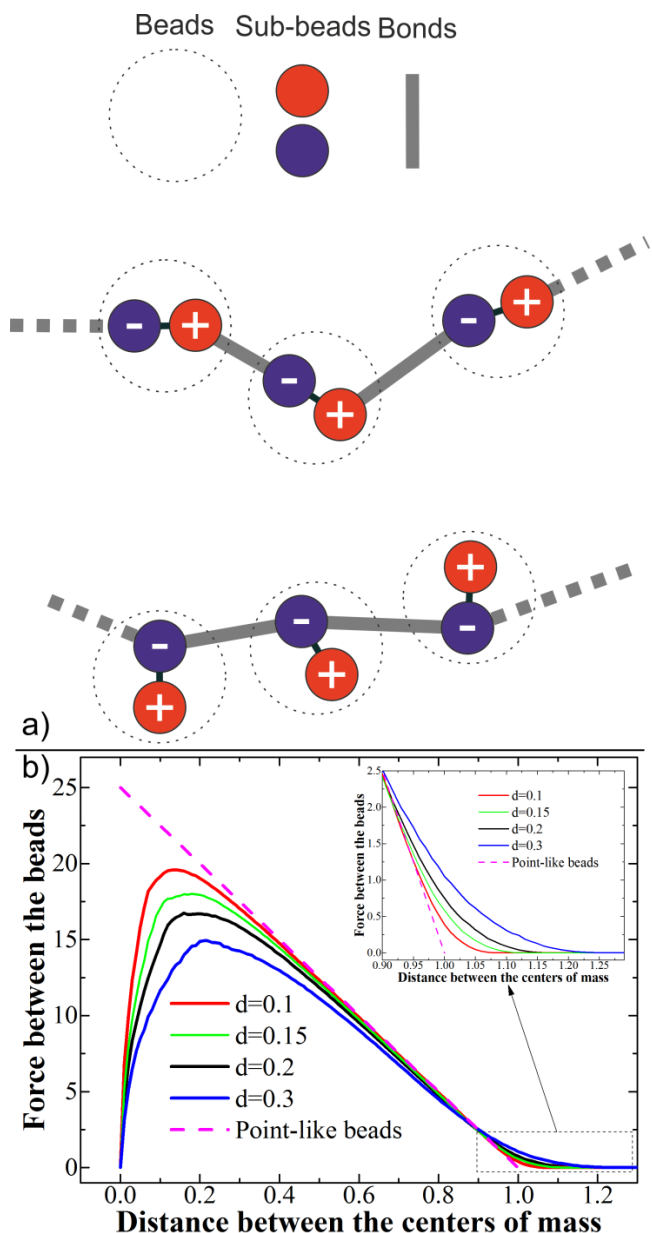

Fig.1 a) Two types of possible positions of the polar beads along the backbone; b) The force between the composite beads for different distances between the sub-beads $d$. The force is averaged over 100000 random orientations at each distance between the centers of mass.

As it was expected, at small enough values of $d$=0.1-0.15 the force between two dumb-bell beads is very close to that between two point-like beads: significant enough differences (>5%) appear only at small distances ($r$<0.33 for $d$=0.15 and $r$<0.2 for $d$=0.1). Smaller values of $d$ would produce even better correspondence; however, further decreasing $d$ would require using smaller timesteps in order for RATTLE to work correctly. At a larger value of $d$=0.2 the difference is more prominent, but the agreement is still reasonable: the deviation rises to 5% at $r$=0.5. The largest studied value of $d$=0.3 seems to be too high as at almost any distance the internal dumb-bell structure significantly influences the interaction between two beads. Therefore, we recommend using $d$=0.1-0.15; the value of $d$=0.2 will be used in further sections for comparison purposes.

Next, we calculated the relation of $\Delta a_{ij}$ between the beads of different types and the Flory-Huggins parameter $\chi$. It is known from the Groot and Warren's work [23] that the relation for the density of $\rho=3$ is rather simple (we consider the simplest case of liquids): $\chi = 0.286 \pm 0.002 \Delta a_{ij}$. Using the same approach as in work [23], we obtained the following expressions for our composite beads for different values of the distance between sub-beads $d$: $\chi = 0.290 \pm 0.003 \Delta a_{ij}$ ($d$=0.1), $\chi = 0.299 \pm 0.004 \Delta a_{ij}$ ($d$=0.15), $\chi = 0.311 \pm 0.004 \Delta a_{ij}$ ($d$=0.2) and $\chi = 0.340 \pm 0.004 \Delta a_{ij}$ ($d$=0.3). We can see that the expressions for $d$=0.1 and 0.15 are rather close to that for the regular point-like beads corroborating that the composite beads at such $d$ values are small enough to be considered as almost point-like; for larger values of $d$, however, the system starts to exhibit the behavior of liquid at a number density higher than 3.

**Liquid bulk permittivity**

In order to understand whether the proposed model of dipoles is applicable to simulate polar liquids, we ran a rather simple test: dielectric liquid with the relative permittivity of $\varepsilon_1>1$ is placed between the flat capacitor plates and the voltage between the plates is measured depending on the dipole moment of the liquid beads. The schematic representation of the system is presented in Fig.2a; the plates are surrounded by non-polar liquid with the relative permittivity $\varepsilon_2=1$ in order to prevent the close interaction of the polar molecules with the charged plates and periodic boundary conditions are used.

The capacitor depicted in Fig.2a can be considered as three capacitors connected in series; the voltage across it can be calculated as $V=V_0(0.25+0.5/\varepsilon_1+0.25)$, where $V_0$ is the voltage across the same capacitor but without the polar liquid. Therefore, $\varepsilon_1=V_0/(2V-V_0)$; the obtained dependence of $\varepsilon_1$ on the dipole moment $qd$ of the polar beads is presented in Fig.2b.

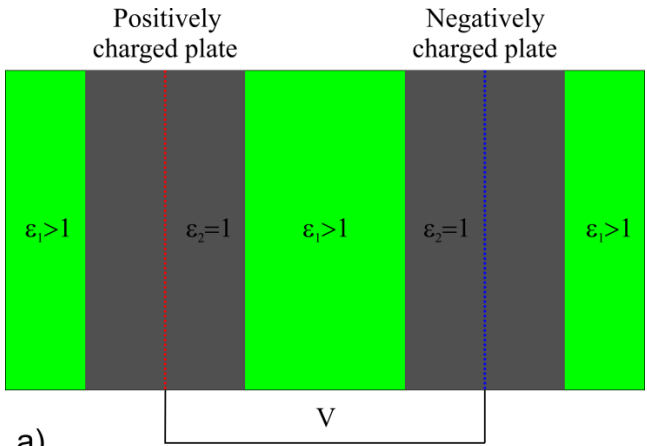

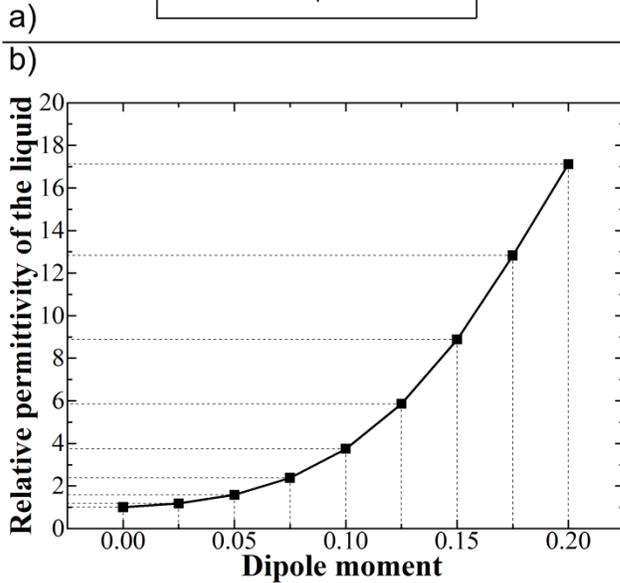

Fig.2 a) Schematic representation of the capacitor used to calculate the bulk relative permittivity of polar liquids. Periodic boundary conditions were applied; b) The obtained dependence of the relative permittivity of polar liquid on the dipole moment of its beads.

We can see that rather high relative permittivity ratios between polar and non-polar liquids can be achieved: for the highest studied dipole moment of 0.7 the dielectric permittivity of the polar liquid was ≈17 times larger than that for the non-polar liquid. This is a rather considerable value: as it was mentioned, we assumed the dielectric permittivity of the media without polar molecules to be equal to 5, so the polar liquid would then have the permittivity of ε=85 which is higher than that of water at room temperature.

Next, we studied the question of the influence of the dipole length *d* on the permittivity of the liquid. As we expected, if the dipole moment was kept constant, no changes within 1% tolerance in the permittivity was found when increasing d from 0.1 to 0.15. When further increasing d to 0.2, the observed permittivities somewhat drop (the observed differences are about 5% for the dipole moments more than 0.6), while at *d*=0.3 the drop was as high as 15%. We assume this is due to the fact that the effective volume of the beads increases (see fig.1b at *r*>1.0) and the reorientation becomes hampered. Therefore, *d*=0.1-0.15 seems to be the safest choice.

## Interaction of charges

Having studied the bulk properties of the polar liquids, let us now investigate the interaction of charges in such liquids. The resulting Coulomb force between two charges in a medium with the dielectric permittivity of ε is equal to $\boldsymbol{F}_{ij} = \dfrac{q_1 q_2 \boldsymbol{r}_{ij}}{4\pi\varepsilon\varepsilon_0 r_{ij}^3}$, i.e. it is ε times smaller than the corresponding force in vacuum. In our treatment of polar liquid, the influence of the medium is treated explicitly and is due to the interaction of the charges with the polar molecules. The latter orient in the electric field around the charges effectively "screening" the interactions between them. Therefore, the question is whether the calculated in the previous values of ε correspond to those observed for the interaction of charges. To that end, two point-like charges were placed into the simulation box filled with polar molecules and the full electrostatic forces acting on the charges were calculated. As it was mentioned, such forces are in some sense the sum of two components: the interaction between the charges themselves and the influence of the polar media. If the forces are averaged over time, the medium contribution can be calculated. It is worth noting that the averaging must be done in the course of the normal system time evolution and not for static charges (i.e. with fixed coordinates) as in the latter case the distribution of polar molecules around charges would be incorrect. Fig. 3 shows the obtained force; the curve was averaged over $150*10^6$ system states.

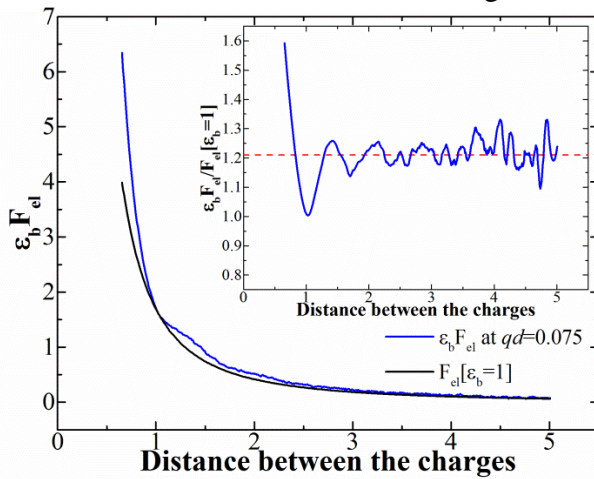

Fig.3 The electrostatic force between two charges in polar liquid normalized by the calculated bulk liquid relative permittivity $\varepsilon_b$ compared to the force in non-polar medium. The inset shows the ratio between these two forces; the curve was smoothed to remove small-scale fluctuations.

We can see that at large enough distances ($r>3$) between charges the observed permittivity levels off and its value is somewhat smaller that that calculated in the bulk (Fig.2). At smaller distances, however, we see an interesting behavior: the observed permittivity starts to fluctuate; when the charges are in direct contact ($r<1$), the permittivity dramatically decreases. This is easily understandable if we consider the fact that each charge is surrounded by solvation shell consisting of oriented polar molecules; at large distances these shells do not feel each other, but when the charges come closer, the shells start to interfere and their structure changes, which causes the observed behavior of the permittivity. The oscillations are due to the fact that the solvation shells consist of several layers of beads and, therefore, their structure is not homogeneous. Such behavior cannot be reproduced by the models in which the polar species are taken into account implicitly via the value of the local permittivity, which may be crucial for reproducing the correct system behavior. The reason for the difference between the permittivity observed for the charges interaction (at large distances, $r>3$) and the permittivity obtained for the bulk liquid is presumably related to the presence of solvation shell as well. The oriented molecules effectively have smaller polarizability compared to the bulk case since the liquid around the charges is already strongly "polarized" in their field, so the electric field of one charge in the vicinity of another charge is reduced not as strong as in the bulk.

**Comparison with the "implicit" method**

In order to demonstrate the necessity of the usage of the explicit treatment of the polar medium, we considered the following system: a hard colloid particle adsorbed at a liquid-liquid interface. This essentially is a model of Pickering emulsions[33]: the hard particle adsorbs at the liquid-liquid interface in order to reduce the number of unfavorable contacts between the immiscible liquids. We assumed one of the liquids to be polar while the other to be non-polar, which is usually the case as one of the liquids in emulsions is water. The colloidal particle was charged: 5% of its units bore a unit charge; the corresponding number of counterions was added to the system to preserve electroneutrality. The overall system size was equal to 72x48x48 (approximately 995000 beads), while the colloidal particle consisted of 9009 beads. Two methods for studying such system were utilized: a) the method developed in this work; b) the method in which the medium permittivity is taken into account implicitly by calculating the local permittivity and subsequent solving the Poisson equation on a mesh described[18]. In the latter case we took the electrostatic smearing radius to be equal to $R_e=0.6$, which allows one to achieve a good (even though not perfect) correspondence with the expression (4) at $D=0.65$[30]. The permittivity of the polar liquid was assumed to be 80, while for the non-polar liquid it was taken equal to 8 (to that end, the $\Gamma$ parameter was taken to be equal to 10). The equilibrium system states for both the methods are presented in Fig.4.

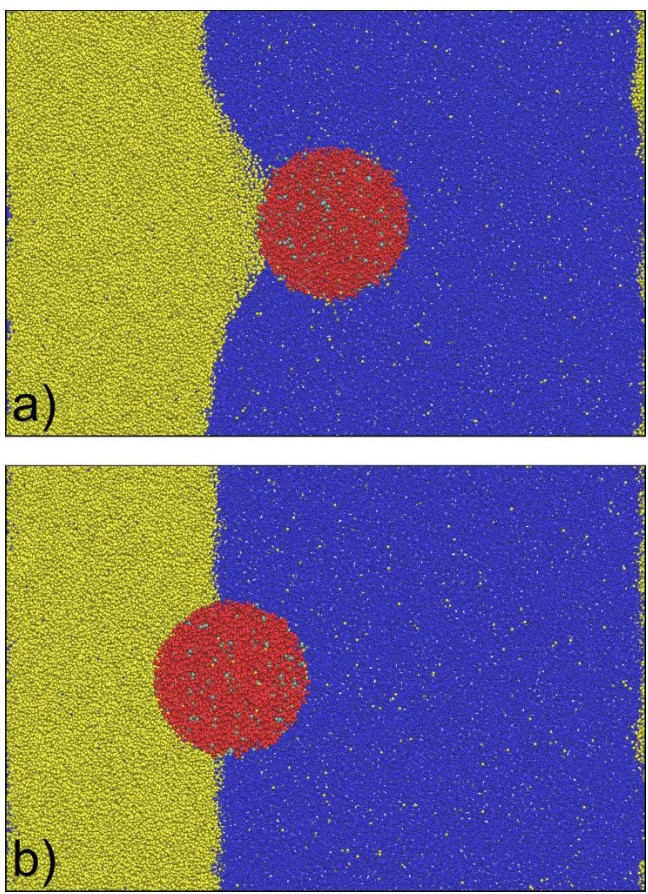

Fig.4 The equilibrium state of a hard colloidal particle at an interface of two liquids with different permittivities calculated using a) the "implicit polarity" model[18]; and b) the model developed in this work. The polar liquid is presented in blue, the non-polar – in yellow, while the colloidal particle is red. The snapshots are generated as cross-sections through the colloidal particle center of mass for better visibility.

We can see that the system state obtained using the model with explicit polarity (developed in this work) is what we expected it to be: the charged colloidal particle remained on the interface but slightly moved into the polar liquid as it is energetically more favorable for the charged molecules to be immersed in a polar medium. The results obtained using the "implicit polarity" model are surprising: the colloidal particle is found on top of a "bump" formed by the molecules of the non-polar liquid. This configuration actually marginally increases the contact area (i.e. the number of unfavorable contacts) between the immiscible liquids (by ~1% in the studied box) instead of significantly decreasing it like in the case of the explicit polarity model. We assume that this happens due to the violation of the Newton's third law: while the charged beads are "dragged" into the polar liquid, there are no electrostatic forces exerted by the charged beads on the liquid beads. We therefore can conclude that such an "implicit" approach should be used with caution when the system contains regions with different permittivities.

**Conclusions**

In this work we developed a method for simulating polar species in the dissipative particle dynamics method. The main idea behind the method is to treat each bead as a dumb-bell (i.e. two sub-beads kept at a fixed distance *d*) instead of a point-like particle; the size of such a dumb-bell should be small enough so that its properties as a whole are not different from a point-like bead. A modified version of the RATTLE routine was used to keep the distance between the sub-beads constant and to preserve the ability to use large timesteps. The sub-beads can bear charges of different sign thus making the bead a dipole; their positions are integrated separately so the rotation of each bead is taken into account automatically. The interaction forces between such composite beads were studied as well as the relation of the interaction parameters between them to the Flory-Huggins theory; it was shown that at *d*=0.1-0.15 the composite beads act essentially as conventional point-like beads. Next, we investigated the relation between the bead dipole moment and the bulk dielectric permittivity; it appeared to be dependent only on the beads dipole moment but not on *d* (if *d* is small enough). The interaction of single charges in polar liquid showed that the observed dielectric permittivity (i.e. the ratio of the force exerted on the charges in vacuum to the corresponding force in the polar liquid) is somewhat smaller than that obtained for the bulk case at large separation between the charges; at distances comparable to the bead size the solvation shells of the charges start to interfere and oscillations in the observed permittivity occur reflecting the presence of the layers of oriented polar molecules surrounding the charges. The latter are presumably the reason for the observed deviations at large distances as well: the beads surrounding the charges are already oriented and therefore effectively have smaller polarizability compared to the bulk liquid, so the field of one charge in the vicinity of another charge is reduced not as strong as in the bulk. This cannot be reproduced within the methods in which the polarity is simulated as dielectric background. Finally, we showed how the violation of the Newton's third law in the latter "implicit polarity" methods results in simulation artifacts. To that end, we investigated the behavior of a charged colloidal particle at a liquid-liquid interface; the permittivity of one of the liquids was 10 times more than that of the second. We obtained that when the polar molecules are treated explicitly, the charged colloidal particle moved into the polar liquid since it is energetically more favorable for the charged molecules to be immersed in a polar medium; however, within the "implicit polarity" method the colloidal particle is found on top of a "bump" formed by the molecules of the non-polar liquid, which increases the interface area between the liquids instead of decreasing it. Such behavior seems to be erroneous.

Concluding, the presented simulation method allows one to deal with polar species without the need of introducing effective local polarizability, which is an assumption with significant drawbacks in the physical behavior. This makes the method a powerful and robust tool that can be applied to various systems in which the polarity of the components can play a crucial role in their behavior.

**Acknowledgments**

We thank prof. I.I.Potemkin and prof. E.Yu.Kramarenko for fruitful discussions. The financial support of the Russian Science Foundation (project 18-73-00128) is greatly acknowledged. The research is carried out using the equipment of the shared research facilities of HPC computing resources at Lomonosov Moscow State University.

**Notes**

The author declares no competing financial interest.

**References**

[1]   I. Nakamura, Z.-G. Wang, Salt-doped block copolymers: ion distribution, domain spacing and effective χ parameter, Soft Matter. 8 (2012) 9356. doi:10.1039/c2sm25606a.

[2]   I. Nakamura, N.P. Balsara, Z.-G. Wang, First-Order Disordered-to-Lamellar Phase Transition in Lithium Salt-Doped Block Copolymers, ACS Macro Lett. 2 (2013) 478–481. doi:10.1021/mz4001404.

[3]   I. Nakamura, Z.-G. Wang, Thermodynamics of Salt-Doped Block Copolymers, ACS Macro Lett. 3 (2014) 708–711. doi:10.1021/mz500301z.

[4]   I. Nakamura, Spinodal Decomposition of a Polymer and Ionic Liquid Mixture: Effects of Electrostatic Interactions and Hydrogen Bonds on Phase Instability, Macromolecules. 49 (2016) 690–699. doi:10.1021/acs.macromol.5b02189.

[5]   I. Nakamura, N.P. Balsara, Z.-G. Wang, Thermodynamics of Ion-Containing Polymer Blends and Block Copolymers, Phys. Rev. Lett. 107 (2011) 198301. doi:10.1103/PhysRevLett.107.198301.

[6]   J. Qin, J.J. De Pablo, Ordering Transition in Salt-Doped Diblock Copolymers, Macromolecules. 49 (2016) 3630–3638. doi:10.1021/acs.macromol.5b02643.

[7]   M.T. Irwin, R.J. Hickey, S. Xie, F.S. Bates, T.P. Lodge, Lithium Salt-Induced Microstructure and Ordering in Diblock Copolymer/Homopolymer Blends, Macromolecules. 49 (2016) 4839–4849. doi:10.1021/acs.macromol.6b00995.

[8]   H.-K. Kwon, B. Ma, M. Olvera de la Cruz, Determining the Regimes of Dielectric Mismatch and Ionic Correlation Effects in Ionomer Blends, Macromolecules. 52 (2019) 535–546. doi:10.1021/acs.macromol.8b02376.

[9]   W. Chu, J. Qin, J.J. de Pablo, Ion Distribution in Microphase-Separated Copolymers with Periodic Dielectric Permittivity, Macromolecules. 51 (2018) 1986–1991. doi:10.1021/acs.macromol.7b02508.

[10]  Z.-G. Wang, Effects of Ion Solvation on the Miscibility of Binary Polymer Blends †, J. Phys. Chem. B. 112 (2008) 16205–16213. doi:10.1021/jp806897t.

[11]  Z.-G. Wang, Fluctuation in electrolyte solutions: The self energy, Phys. Rev. E. 81 (2010) 021501. doi:10.1103/PhysRevE.81.021501.

[12]  A.M. Rumyantsev, E.Y. Kramarenko, Two regions of microphase separation in ion-containing polymer solutions, Soft Matter. 13 (2017) 6831–6844. doi:10.1039/C7SM01340J.

[13]  J.H. Choi, Y. Ye, Y.A. Elabd, K.I. Winey, Network structure and strong microphase separation for high ion conductivity in polymerized ionic liquid block copolymers, Macromolecules. 46 (2013) 5290–5300. doi:10.1021/ma400562a.

[14]  Y. Ye, S. Sharick, E.M. Davis, K.I. Winey, Y.A. Elabd, High hydroxide conductivity in polymerized ionic liquid block copolymers, ACS Macro Lett. 2 (2013) 575–580. doi:10.1021/mz400210a.

[15]  S. Tyagi, M. Süzen, M. Sega, M. Barbosa, S.S. Kantorovich, C. Holm, An iterative, fast, linear-scaling method for computing induced charges on arbitrary dielectric boundaries, J.

Chem. Phys. 132 (2010) 154112. doi:10.1063/1.3376011.
[16] V. Jadhao, F.J. Solis, M.O. de la Cruz, Simulation of Charged Systems in Heterogeneous Dielectric Media via a True Energy Functional, Phys. Rev. Lett. 109 (2012) 223905. doi:10.1103/PhysRevLett.109.223905.
[17] V. Jadhao, F.J. Solis, M. Olvera De La Cruz, A variational formulation of electrostatics in a medium with spatially varying dielectric permittivity, J. Chem. Phys. 138 (2013). doi:10.1063/1.4789955.
[18] R.D. Groot, Electrostatic interactions in dissipative particle dynamics—simulation of polyelectrolytes and anionic surfactants, J. Chem. Phys. 118 (2003) 11265. doi:10.1063/1.1574800.
[19] F. Fahrenberger, O.A. Hickey, J. Smiatek, C. Holm, The influence of charged-induced variations in the local permittivity on the static and dynamic properties of polyelectrolyte solutions, J. Chem. Phys. 143 (2015). doi:10.1063/1.4936666.
[20] P.J. Hoogerbrugge, J.M.V.A. Koelman, Simulating Microscopic Hydrodynamic Phenomena with Dissipative Particle Dynamics, Europhys. Lett. 19 (1992) 155–160. doi:10.1209/0295-5075/19/3/001.
[21] A.G. Schlijper, P.J. Hoogerbrugge, C.W. Manke, Computer simulation of dilute polymer solutions with the dissipative particle dynamics method, J. Rheol. 39 (1995) 567–579. doi:10.1122/1.550713.
[22] P. Español, P. Warren, Statistical Mechanics of Dissipative Particle Dynamics, Europhys. Lett. 30 (1995) 191–196. doi:10.1209/0295-5075/30/4/001.
[23] R.D. Groot, P.B. Warren, Dissipative particle dynamics: Bridging the gap between atomistic and mesoscopic simulation, J. Chem. Phys. 107 (1997) 4423–4435. doi:10.1063/1.474784.
[24] J. Guo, H. Liang, Z.-G. Wang, Coil-to-globule transition by dissipative particle dynamics simulation, J. Chem. Phys. 134 (2011) 244904. doi:10.1063/1.3604812.
[25] R.D. Groot, T.J. Madden, Dynamic simulation of diblock copolymer microphase separation, J. Chem. Phys. 108 (1998) 8713. doi:10.1063/1.476300.
[26] Y. Li, H.-J. Qian, Z.-Y. Lu, The influence of one block polydispersity on phase separation of diblock copolymers: The molecular mechanism for domain spacing expansion, Polymer. 54 (2013) 3716–3722. doi:10.1016/j.polymer.2013.04.064.
[27] G. Raos, M. Casalegno, Nonequilibrium simulations of filled polymer networks: searching for the origins of reinforcement and nonlinearity., J. Chem. Phys. 134 (2011) 054902. doi:10.1063/1.3537971.
[28] A.A. Gavrilov, A. V. Chertovich, P.G. Khalatur, A.R. Khokhlov, Study of the Mechanisms of Filler Reinforcement in Elastomer Nanocomposites, Macromolecules. 47 (2014) 5400–5408. doi:10.1021/ma500947g.
[29] A.A. Gavrilov, P. V. Komarov, P.G. Khalatur, Thermal Properties and Topology of Epoxy Networks: A Multiscale Simulation Methodology, Macromolecules. 48 (2015) 206–212. doi:10.1021/ma502220k.
[30] A.A. Gavrilov, A. V. Chertovich, E.Y. Kramarenko, Dissipative particle dynamics for systems with high density of charges: Implementation of electrostatic interactions, J. Chem. Phys. 145 (2016) 174101. doi:10.1063/1.4966149.
[31] M.P. Allen, D.J. Tildesley, Computer simulation of liquids, Clarendon Press, Oxford, UK, 1991.
[32] H.C. Andersen, Rattle: A "velocity" version of the shake algorithm for molecular dynamics calculations, J. Comput. Phys. 52 (1983) 24–34. doi:10.1016/0021-9991(83)90014-1.
[33] S.U. Pickering, CXCVI.—Emulsions, J. Chem. Soc., Trans. 91 (1907) 2001–2021. doi:10.1039/CT9079102001.